%% file: eprint.tex
\documentclass[10pt]{article}
\usepackage{graphicx}

\def\Title#1{\begin{center} {\Large #1} \end{center}}
\def\Author#1{\begin{center}{ \sc #1} \end{center}}
\def\Address#1{\begin{center}{ \it #1} \end{center}}

\newcommand\pubblock{\rightline{\begin{tabular}{l} Proceedings of the Second Annual LHCP\\ \pubnumber\\
         \pubdate  \end{tabular}}}

\newenvironment{Abstract}{\begin{quotation} \begin{center} 
             \large ABSTRACT \end{center}\bigskip 
      \begin{center}\begin{large}}{\end{large}\end{center} \end{quotation}}

\newenvironment{Presented}{\begin{quotation} \begin{center} 
             PRESENTED AT\end{center}\bigskip 
      \begin{center}\begin{large}}{\end{large}\end{center} \end{quotation}}

\input econfmacros.tex

\usepackage{color}

\newcommand\pubnumber{ CMS CR-2014/181 }

\newcommand\pubdate{\today}

\def\affiliation{
On behalf of the CMS Experiment, \\
Department of Physics \\
Universit\'{e} Libre de Bruxelles, Universit\'{a} di Roma La Sapienza }

\begin{document}

\large
\begin{titlepage}
\pubblock

\vfill
\Title{  $ZZ$ cross section measurement and limits on anomalous Triple Gauge Couplings at CMS experiment (preliminary results)  }
\vfill

\Author{ Luca Perni\'{e} on behalf of the CMS Collaboration}
\Address{\affiliation}
\vfill
\begin{Abstract}
The measurement of the ZZ production cross section in
proton-proton collisions at center-of-mass energies of 7 and 8 TeV
from data acquired by the CMS experiment at the LHC is presented.
The decay channel $ZZ \rightarrow 2l2\nu$ is selected
from data corresponding to an integrated luminosity of 5.1 fb$^{-1}$
at 7 TeV and 19.6 fb$^{-1}$ at 8 TeV.\\
The measured cross sections,
$\sigma({\rm pp} \rightarrow ZZ) = 5.0_{-1.4}^{+1.5}\,({\rm
stat})\,_{-1.0}^{+1.3}\,({\rm syst})\,\pm 0.2\,({\rm lumi})\;{\rm
pb}$ at 7 TeV, and
$\sigma({\rm pp} \rightarrow ZZ) = 6.8_{-0.8}^{+0.8}\,({\rm
stat})\,_{-1.4}^{+1.8}\,({\rm syst})\,\pm 0.3\,({\rm lumi})\;{\rm
pb}$ at 8 TeV,
are in good agreement with the standard model NLO predictions. The selected data are analyzed to search for
anomalous triple gauge couplings involving the $ZZ$ final state.
\end{Abstract}
\vfill

\begin{Presented}
The Second Annual Conference\\
 on Large Hadron Collider Physics \\
Columbia University, New York, U.S.A \\ 
June 2-7, 2014
\end{Presented}
\vfill
\end{titlepage}
\def\thefootnote{\fnsymbol{footnote}}
\setcounter{footnote}{0}
%

\normalsize 


\section{Introduction}
The measurement of the $Z$ bosons pair production into the $2l2\nu$ final state cross section at 7 and 8 TeV at CMS~\cite{CMSa} experiment is presented.
In the Standard Model (SM), a $Z$ bosons pair is produced through non-resonant processes or through the decay of a Higgs boson, and the predictions
of the SM for this rare electroweak process can be tested with precision for the first time at the LHC.\\
Vector bosons are expected to couple in triplets (e.g. $WWZ$) and in quartets (e.g. $WWZZ$) as
a consequence of the non-abelian structure of the electroweak gauge theory. All couplings involving
the bosons with neutral electric charge are expected to vanish at tree level leading to
the absence of triple couplings for $Z\gamma\gamma$, $ZZ\gamma$ and $ZZZ$. An observation of these anomalous
couplings would therefore be an indication of new physics.
This new physics can be parametrized as anomalous coupling
constants (aTGC) for neutral triple gauge boson interactions $\gamma ZZ$ and $ZZZ$,
and it can appear as deviations from the SM, e.g. in the $Z$ boson transverse momentum ($p_T$) spectrum.\\
The $ZZ$ production cross section is expected to be $6.46^{+4.7\%}_{-3.3\%}$ pb at
$\sqrt{s}$ = 7 TeV and $7.92^{+4.7\%}_{-3.0\%}$ pb at $\sqrt{s}$  = 8 TeV at next-to-leading order (NLO)~\cite{Campbell}.
At the tree level, $ZZ$ final states are primarily produced in the t- and u-channels and gluon-gluon
fusion. NLO electroweak correction calculations have been recently published~\cite{Tobias} and will be applied in next approved results.\\
Data corresponding to an integrated luminosity of about 5.1 fb$^{-1}$ at 7 TeV and 19.5 fb$^{-1}$ at 8 TeV have been analyzed
and the cross-sections measurement is in good agreement with the SM NLO predictions.

\section{Event selection}
The $2l2\nu$ final state is characterized from the presence of 2 energetic leptons ($e$/$\mu$) and high transverse missing energy ($E_t^{miss}$).\\
To better describe the presence of neutrinos, a new variable, called $reduced$ $E_t^{miss}$, has been defined as done successfully in D0 experiment~\cite{D0}.
The general concept behind a reduced $E_t^{miss}$ is to reduce the instrumental contribution
by considering possible contributions to fake missing transverse energy on an
event-by-event basis. In each event, $E_t^{miss}$ and jets momentum are decomposed along an orthogonal set
of axes in the transverse plane of the detector. One of the axes is defined by the transverse
momentum of the charged dilepton system, the other perpendicular to it.\\
We define the recoil of the dilepton system in two different ways: the clustered recoil ($\vec{R}_{clust}$),
the vectorial sum of the jets reconstructed in the event, and the unclustered recoil ($\vec{R}_{uncl}$),
the sum of all the candidates in the event, with the exception of the two leptons. On each
axis $i$, the reduced-$E_t^{miss}$ projection is defined as:
$$reduced-E_t^{miss^i} = -q^i_T-R^i_{c-u}$$
where $R^i_{c-u} $ is $\vec{R}_{clust}$ or $\vec{R}_{uncl}$ depending on which minimize the absolute value of the $reduced-E_t^{miss}$ component.
The selection chosen to remove the presence of background is reported in table~\ref{tab:selectioncuts}.
The cuts have been chosen, among all the possible combinations, to yield the lowest statistical and systematic uncertainty.
The selection is identical for 7 and 8 TeV data and for the search of anomalous couplings.

\begin{table}[h]
 \begin{center}
 \caption{Summary of the optimal signal selection.}
 \label{tab:selectioncuts}
 \begin{tabular}{cl}
   \hline
   Variable  & Value \\
   \hline
   \hline
   Dilepton invariant mass & $|m(ll)-91| < 7.5$ GeV \\
   Dilepton $p_T$            & $q_T > 45$ GeV \\
   b-tag veto                & based on vertex info. (for jet with $p_T > 20$ GeV) \\
   Jet veto                  & no jets with $p_T > 30\, GeV$\\
   Reduced $E_t^{miss}$              & $> 65$ GeV \\
   $E_t^{miss}$ balance              & $0.4 < B < 1.8$\\
   $\Delta\phi$($E_t^{miss}$,jet)    & $> 0.5$ rad \\
   $\Delta\phi$($E_t^{miss}$,lept.)  & $> 0.2$ rad \\
   Lepton veto               & no additional leptons ($e$/$\mu$) with $p_T>10/3 GeV$ \\
   \hline
 \end{tabular}
 \end{center}
\end{table}

\section{Backgrounds estimation}
\begin{figure}[htb]
\centering
\includegraphics[height=3in]{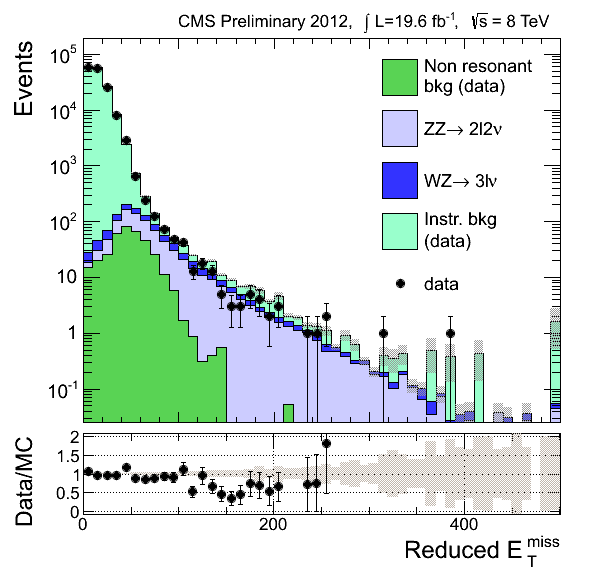}
\caption{Reduced $E_t^{miss}$ spectrum in the inclusive ll channel at 8 TeV, using the DY-template derived from the photon sample at preselection level.}
\label{fig:figure1}
\end{figure}

While the $WZ$ background is estimated from simulation, after cross-checking that the simulation agree with the data in a control region
with three identified leptons, a data-driven method is applied to estimate the total number of background events from
processes which do not involve a $Z$ boson: i.e. $WW$ and top production~\cite{CMS}.
In order to measure this contribution, a control sample
based on $e\mu$ candidate events is selected by applying the same selection cuts as in the main
analysis. The yields in the same-flavour channels ($ee$ and $\mu\mu$) are then obtained by scaling
the number of events in the control sample. The re-scaling is done by means of data-driven
factors, measured from the side-bands of the $Z$ mass peak region (55-70 GeV/c$^2$ and 110-200 GeV/c$^2$).\\
The estimation of the Drell-Yan (DY) background requires special care. Although the DY process does not include genuine
$E_t^{miss}$ from neutrinos, the tail of the reduced-$E_t^{miss}$ distribution
can be contaminated by these events due to detector energy resolution, jet energy mis-measurements,
pile-up energy fluctuations, and instrumental noise. 
DY it has been estimated using a process which is topologically similar, with a much higher cross section, as the production of
prompt isolated photons in association with jets (i.e. $\gamma$+jets).  
The DY data driven estimation uses the followings steps:
\begin{itemize}
\item The kinematic selection to remove processes with real neutrinos, selecting a high purity photon sample in data
\item The application of weights to the $\gamma~p_T$ spectrum to match the $Z$ $p_T$ spectrum from the dilepton sample
\item The application of the full selection to the weighted photon sample
\item The subtraction of the the EWK $\gamma$+$E_t^{miss}$ events using the simulated predictions
\end{itemize}
The remaining contribution has been used as DY estimation (light green component of Fig.~\ref{fig:figure1}) 

\section{Results}
\begin{figure}[htb]
\centering
\includegraphics[height=3in]{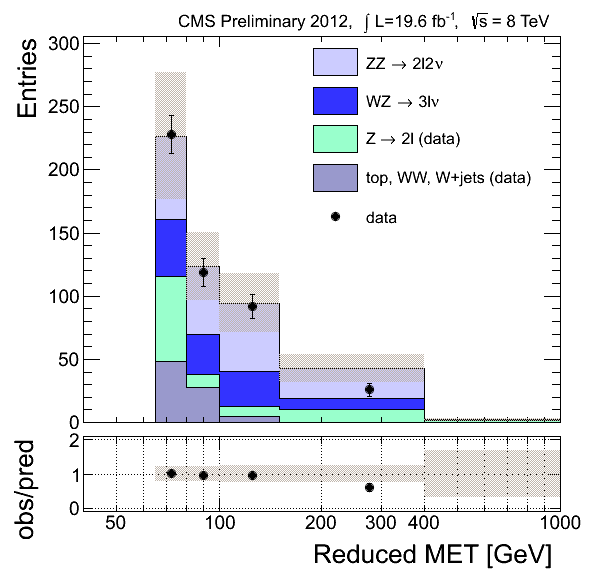}
\caption{Reduced $E_t^{miss}$ distribution in ll channels, after the full selection, at 8 TeV. DY and non-resonant backgrounds are estimated with data-driven methods. The grey error band includes statistical and systematic uncertainties on the predicted yields. In the bottom plots, error bars and bands are relative to the total predicted yields.}
\label{fig:figure2}
\end{figure}

\begin{figure}[htb]
\centering
\includegraphics[height=3in]{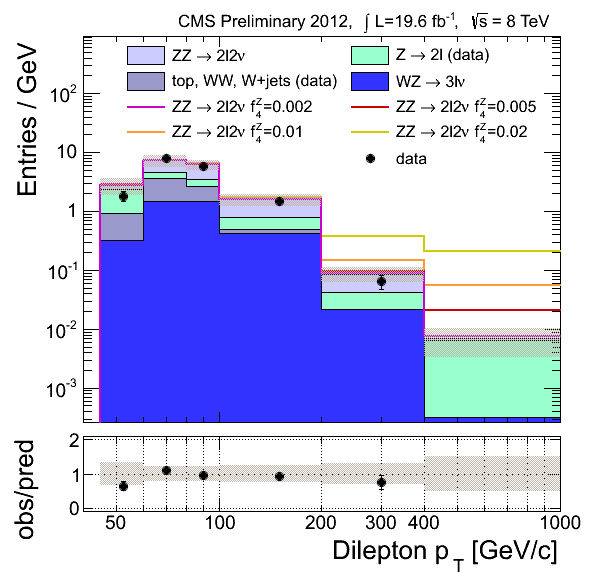}
\caption{Dilepton transverse momentum distributions at 8 TeV. DY and non-resonant backgrounds are estimated with data-driven methods. The grey error band includes statistical and systematic uncertainties on the predicted yields. In the bottom plots, error bars and bands are relative to the total predicted yields.}
\label{fig:figure3}
\end{figure}

We extract the $ZZ$ production cross section using a profile likelihood fit to the reduced-$E_T^{miss}$
distribution (Fig.~\ref{fig:figure2}), which takes into account the expectations for the different background
processes and the $ZZ$ signal. Each systematic uncertainty is introduced to the fit as a nuisance
parameter with log-normal prior. For the signal we consider a further multiplicative factor,
 which is the ratio of the cross section to be measured in data to the expected theoretical value.
Maximizing the profile likelihood, we obtain the $ZZ$ production
cross section from the signal strength fit, as well as optimal fits of the background yields from a fine adjustment of the nuisance parameters.
\\
The measured cross sections are:
$$ \sigma({\rm pp} \rightarrow ZZ) = 5.0_{-1.4}^{+1.5}\,({\rm 
stat})\,_{-1.0}^{+1.3}\,({\rm syst})\,\pm 0.2\,({\rm lumi})\;{\rm pb}$$
at 7 TeV, and
$$\sigma({\rm pp} \rightarrow ZZ) = 6.8_{-0.8}^{+0.8}\,({\rm
stat})\,_{-1.4}^{+1.8}\,({\rm syst})\,\pm 0.3\,({\rm lumi})\;{\rm pb}$$
at 8 TeV, in good agreement with the SM NLO predictions.
Since no deviation from SM have been observed, we search for deviations in the kinematics due to the presence of aTGCs.\\
Neutral couplings V$^{(\star)}$ZZ (V = Z, $\gamma$) can be described
using the following effective Lagrangian:
$${\cal L}_{\rm VZZ} \; = \; -\frac{e}{M_{\rm Z}^2} \left\{
\left[f_4^\gamma\left(\partial_\mu F^{\mu\alpha}\right)
+f_4^Z\left(\partial_\mu Z^{\mu\alpha}\right)\right]
Z_\beta\left(\partial^\beta Z_\alpha\right)
-\left[f_5^\gamma\left(\partial^\mu F_{\mu\alpha}\right)
+f_5^Z\left(\partial^\mu Z_{\mu\alpha}\right)\right]
\tilde{Z}^{\alpha\beta}Z_\beta
\right\}$$
Coefficients $f_i^\gamma$ and $f_i^Z$ correspond to couplings $\gamma^{(\star)}$ZZ and Z$^{(\star)}$ZZ, respectively. All the operators
are Lorentz-invariant and ${\rm U}(1)_{EM}$ gauge-invariant, but not invariant under
${\rm SU}(2)_L\times {\rm U}(1)_Y$ gauge symmetry. The terms corresponding to $f_4^{\rm V}$ parameters violate the CP
symmetry, while the terms corresponding to $f_5^{\rm V}$ parameters conserve CP.\\
Figure~\ref{fig:figure3} shows the charged dilepton $p_T$ distribution after the full selection,
in data and MC, including SHERPA~\cite{SHERPA} samples with different values of the aTGC parameters.
\\
In the search for aTGCs, the SM production of $ZZ$ represents a background, while the sole contribution of the aTGCs constitutes the signal. This signal is built from the SHERPA, by subtracting the SM contribution to the charged dilepton $p_T$.\\
The limits are calculated with a profile likelihood method. We set one-dimensional limits
on the four parameters, i.e. varying independently a single parameter at a time, while fixing
the other three to 0. The 95\% C.L. one-dimensional limits on the four parameters are reported
in Table~\ref{tab:limits_all} for 7 TeV and 8 TeV.
\begin{table}[h]
\begin{center}
\caption{Summary of 95\% C.L. intervals for the neutral aTGC coefficients, set by this analysis using the 7 and 8 TeV CMS
  datasets. The expected 95\% C.L. intervals obtained using the 7 and 8 TeV simulated samples are also shown. No form factor is used.}
\label{tab:limits_all}
\small
\begin{tabular}{c|cccc}
  \hline
  Dataset & $f_4^Z$ & $f_4^\gamma$ & $f_5^Z$ & $f_5^\gamma$ \\
  \hline
  \hline
  7 TeV    & [-0.0088; 0.0085] & [-0.0098; 0.011]  & [-0.0096; 0.0096] & [-0.011 ; 0.010] \\
  \hline
  8 TeV    & [-0.0038; 0.0040] & [-0.0049; 0.0039] & [-0.0041; 0.0038] & [-0.0049; 0.0046]\\
  \hline
  Combined & [-0.0030; 0.0034] & [-0.0039; 0.0031] & [-0.0036; 0.0032] & [-0.0038; 0.0038]\\
  \hline
\end{tabular}
\end{center}
\end{table}

\section{Conclusions}
We have measured the $ZZ$ production cross section from proton-proton collisions at center-of-mass energies of 7 and 8 TeV.
The data sample selected for our study corresponds to about 5.1 fb$^{-1}$ of
integrated luminosity at 7 TeV, and about 19.6 fb$^{-1}$ at 8 TeV.
The measured cross section is in good agreement with the SM NLO predictions,
and the selected data were also analyzed to search for anomalous triple gauge
couplings involving the ZZ final state. In the absence of signs of new physics, we have set limits on the relevant aTGC parameters.

\newpage

\end{document}

%% file: econfmacros.tex
\def\beq{\begin{equation}}
\def\eeq#1{\label{#1}\end{equation}}
\def\eeqn{\end{equation}}

\def\beqa{\begin{eqnarray}}
\def\eeqa#1{\label{#1}\end{eqnarray}}
\def\eeqan{\end{eqnarray}}

\let\bar=\overbar

\def\Dslash{\not{\hbox{\kern-4pt $D$}}}
\def\dslash{\not{\hbox{\kern-2pt $\del$}}}

\def\msb{{\bar{\ssstyle M \kern -1pt S}}}

